\documentclass[sigconf]{acmart}

\AtBeginDocument{%
  \providecommand\BibTeX{{%
    \normalfont B\kern-0.5em{\scshape i\kern-0.25em b}\kern-0.8em\TeX}}}

\copyrightyear{2024}
\acmYear{2024}
\setcopyright{acmlicensed}\acmConference[WWW '24]{Proceedings of the ACM
Web Conference 2024}{May 13--17, 2024}{Singapore, Singapore}
\acmBooktitle{Proceedings of the ACM Web Conference 2024 (WWW '24), May
13--17, 2024, Singapore, Singapore}
\acmDOI{10.1145/3589334.3648151}
\acmISBN{979-8-4007-0171-9/24/05}

\usepackage{algorithm}
\usepackage{algorithmic}
\usepackage{appendix}
\usepackage{booktabs}
\usepackage{amsmath}
\usepackage{multicol}
\usepackage{multirow}
\usepackage{array}
\usepackage{bm}
\usepackage{pbox}
\usepackage{makecell}
\newcolumntype{P}[1]{>{\centering\arraybackslash}p{#1}}

\usepackage{xcolor}

\usepackage[belowskip=0pt,aboveskip=0pt]{caption}

\acmConference[Conference acronym 'XX]{Make sure to enter the correct
  conference title from your rights confirmation emai}{June 03--05,
  2018}{Woodstock, NY}
\acmPrice{15.00}
\acmISBN{978-1-4503-XXXX-X/18/06}

\begin{document}


\title{MemeCraft: Contextual and Stance-Driven \\Multimodal Meme Generation}



\author{Han Wang}
\orcid{0009-0007-4486-0693}
\affiliation{%
  \department{Information Systems Technology and Design}
  \institution{Singapore University of Technology and Design}
  \streetaddress{8 Somapah Road}
  \country{Singapore}
  \postcode{487372}
}
\email{han_wang@sutd.edu.sg}

\author{Roy Ka-Wei Lee}
\orcid{0000-0002-1986-7750}
\affiliation{%
  \department{Information Systems Technology and Design}
  \institution{Singapore University of Technology and Design}
  \streetaddress{8 Somapah Road}
  \country{Singapore}
  \postcode{487372}
}
\email{roy_lee@sutd.edu.sg}




\begin{abstract}
Online memes have emerged as powerful digital cultural artifacts in the age of social media, offering not only humor but also platforms for political discourse, social critique, and information dissemination. Their extensive reach and influence in shaping online communities' sentiments make them invaluable tools for campaigning and promoting ideologies. 
Despite the development of several meme-generation tools, there remains a gap in their systematic evaluation and their ability to effectively communicate ideologies.  Addressing this, we introduce \textsf{MemeCraft}, an innovative meme generator that leverages large language models (LLMs) and visual language models (VLMs) to produce memes advocating specific social movements. \textsf{MemeCraft} presents an end-to-end pipeline, transforming user prompts into compelling multimodal memes without manual intervention. Conscious of the misuse potential in creating divisive content, an intrinsic safety mechanism is embedded to curb hateful meme production. 
Our assessment, focusing on two UN Sustainable Development Goals—Climate Action and Gender Equality—shows \textsf{MemeCraft}'s prowess in creating memes that are both funny and supportive of advocacy goals. This paper highlights how generative AI can promote social good and pioneers the use of LLMs and VLMs in meme generation.

\end{abstract}

\ccsdesc[500]{Do Not Use This Code~Generate the Correct Terms for Your Paper}
\ccsdesc[300]{Do Not Use This Code~Generate the Correct Terms for Your Paper}
\ccsdesc{Do Not Use This Code~Generate the Correct Terms for Your Paper}
\ccsdesc[100]{Do Not Use This Code~Generate the Correct Terms for Your Paper}

\keywords{mutlimodal memes, meme generation, visual language model}

\maketitle

\section{Introduction}
\begin{figure}[t]
  \centering
  \includegraphics[width=0.4\textwidth]{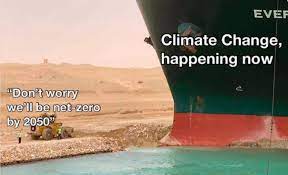}
  \caption{An example meme campaigning for climate change.}
  \label{fig:climate_change_online_memes}
\end{figure}

Online memes represent digital cultural artifacts that have become pervasive in the era of social media. Typically comprised of images, videos, or text, memes frequently convey humor, irony, or satire, rendering them highly shareable and relatable to a broad audience. These memes can disseminate rapidly, at times achieving viral status, and wield substantial influence in shaping the collective sentiments and trends of the online community. Beyond their comedic aspect, memes can also serve as conduits for political discourse, social critique, and the dissemination of information. Multiple research endeavors have explored the communicative prowess of memes, suggesting their efficacy as both a communication and campaigning instrument~\cite{msugheter2020internet, moody2019analysis, kulkarni2017internet}. Such studies affirm memes' unparalleled reach and persuasiveness, often harnessed by individuals and groups to campaign for social movements or advancing ideologies. For instance, \cite{zhangpinto2021} found that exposure to climate change-themed memes increases individuals' intentions to engage in online civic discourse concerning climate change. 
Figure~\ref{fig:climate_change_online_memes} shows an exmaple of a memes  advocating for climate change. 

In response to the soaring popularity and applicability of memes, both academic researchers and the tech industry have presented an array of tools to aid in meme generation~\cite{peirson2018dank,sadasivam2020memebot,vyalla2020memeify}. Semi-automated tools such as Imgflip\footnote{\url{https://imgflip.com/}} offer users the capacity to overlay text on pre-existing meme templates. 

While these tools enhance meme creation efficiency through deep learning, they typically require extensive training with large datasets. For example, Dank Learning~\cite{peirson2018dank} used a dataset of 400,000 memes from \cite{memegenerator2018}, and Memeify~\cite{vyalla2020memeify} collected 1.1 million memes from \cite{quickmeme2016} for training. In contrast, our approach leverages large models effectively with only a few demonstration memes, achieving similar performance while drastically reducing training costs.
Moreover, current evaluations of meme-generation tools are not systematic. Existing studies primarily assess the resemblance of automatically generated memes to their manually crafted counterparts~\cite{peirson2018dank,sadasivam2020memebot}. They often overlook evaluating the ability of these generated memes to effectively and persuasively communicate specific ideologies.

To address these gaps, we introduce \textsf{MemeCraft}, a state-of-the-art, multimodal meme generator that harnesses the capabilities of large language models (LLMs) and visual language models (VLMs) to generate humorous memes that convey specific contexts and stances as directed by users. Specifically, \textsf{MemeCraft} is an end-to-end pipeline that generates multimodal memes from specially designed prompts that aim to advocate specific social movements. Our proposed meme generator automatically takes an image and generates a text overlay to convert the image into a humorous meme that advocates the specified social causes. With \textsf{MemeCraft}, users can effortlessly craft natural, compelling memes, aligning with extensive social and communication campaigns.

Recent studies have also indicated that malicious users have created and circulated hateful memes, exacerbating societal discord, perpetuating stereotypes, and disseminating false information~\cite{hermida2023detecting, sharmadetecting}. To counteract potential misuse of \textsf{MemeCraft}, we have implemented a self-regulating safety mechanism. This feature ensures \textsf{MemeCraft} does not produce hateful memes, maintaining a platform free from abusive or offensive content.

To evaluate the effectiveness of \textsf{MemeCraft} as a social campaigning tool, we first generate a large dataset of memes concentrated around two of the United Nations Sustainable Development Goals (UN SDGs)~\cite{hak2016sustainable}: ``\textit{Climate Action}'' and ``\textit{Gender Equality}''. The dataset encompasses memes both in favor of and against these pivotal goals. Subsequently, we conduct a comprehensive human evaluation to assess the quality of these generated memes on four primary factors: \textit{authenticity}, \textit{hilarity}, \textit{message conveyance}, and \textit{persuasiveness}. Additionally, we also evaluate the robustness of \textsf{MemeCraft}'s safety mechanism against the creation of hateful content.

Our contributions can be summarized as follows: (i) We introduced \textsf{MemeCraft}\footnote{\url{https://github.com/Social-AI-Studio/MemeCraft}} which is an innovative meme generator utilizing LLMs or VLMs to produce large-scale advocacy memes. To our knowledge, this is the pioneering exploration of adapting LLMs or VLMs for meme generation. (ii) We conducted an extensive human evaluation to assess the effectiveness of our proposed meme generator in promoting SDGs (i.e., climate action and gender equality). Our experiment results show that \textsf{MemeCraft} outperformed the state-of-the-art meme generators in generating humorous and persuasive memes that advocate these social movements. At a higher level, our study explores and demonstrates the potential of leveraging generative AI to promote social good.

\section{Related Works}

\subsection{Meme Anaylsis and Generation}


Internet memes are a significant cultural phenomenon in online communication. Extensive research has focused on analyzing topics \cite{du2020understanding}, semantics \cite{xu2022met}, and emotions \cite{sharma2020semeval} conveyed in memes. An important application of meme analysis is the detection of hateful memes. Detection of hateful memes is crucial due to their potential misuse for spreading harmful messages \cite{kiela2020hateful, mathias2021fhmfg}, misinformation \cite{pramanick2021harmemes, naseem2023multimodal}, and propaganda \cite{dimitrov2021detecting}. Efforts to develop models for detecting harmful memes have intensified in academia and industry \cite{pramanick2021momenta, thakur2022multimodal, kirk2021memes, lee2021disentangling, hee2022explaining, cao2023pro, hee2023decoding, zhu2022multimodal, cao2022prompting}.


The generation of memes has evolved significantly over time. Initial models primarily relied on rule-based methods for meme creation~ \cite{wang2015icanhascheezburger,oliveira2016automatic}. With the emergence of deep learning techniques, novel methodologies have been proposed. Peirson and Tolunay~\cite{peirson2018dank} presented the Dank Learning approach, which employs the Inception V3 architecture for image encoding and an attention-driven, deeper-layer LSTM for decoding, facilitating the generation of meme captions for images. Subsequent models, including \textit{MemeBot}~\cite{sadasivam2020memebot} and \textit{Memeify}~\cite{vyalla2020memeify}, incorporate text input and harness transformer-based architectures. Evaluation metrics for these models commonly include differentiability between generated memes and human-craft memes, user satisfaction, and the humor quotient.


Collectively, prior research has provided valuable insights into the methodologies of meme generation. Building on these foundational works, our research introduces a novel meme generator designed to bolster social movements. Differing from existing tools, our proposed model employs LLMs and VLMs to craft humorous memes at scale to communicate specific social causes and stances as determined by the users. Additionally, we enhance the existing studies by conducting a thorough human-centric evaluation, assessing the efficacy of our model in advocating social campaigns.


\subsection{Large Language and Visual Language Models}

Large language models (LLMs), characterized by vast parameter sizes often in the range of billions, are pre-trained on expansive natural language datasets. Notable examples include GPT-3~\cite{brown2020language}, T5~\cite{raffel2020exploring}, and BLOOM~\cite{scao2022bloom}. A defining feature of these LLMs is their ability to perform exceptionally well on novel tasks with only task instructions provided, a capability termed as zero-shot in-context learning. To enhance this performance, certain models undergo post-pre-training refinements. Specifically, T0~\cite{sanh2022multitask} and FLAN~\cite{wei2021finetuned} are fine-tuned across various tasks, outperforming GPT-3 in zero-shot scenarios. InstructGPT\cite{ouyang2022training} undergoes refinement from GPT-3 via reinforcement learning from human feedback (RLHF), thereby improving its adherence to instructions. Additionally, ChatGPT\footnote{\url{https://openai.com/blog/chatgpt}}, which originates from InstructGPT, is fine-tuned with conversational datasets using RLHF, enabling it to engage in dialogues and provide detailed responses to user queries. Touvron et al. \cite{touvron2023llama} introduced LLaMA, a suite of efficient Large Language Models with parameter sizes ranging from 7B to 65B. Despite their reduced scale, these LLMs have proven to be highly competitive when compared to their larger counterparts. In this study, we investigated the use of online LLM APIs, such as ChatGPT, as well as offline models like LLaMA for meme generation. To our knowledge, this is the inaugural study that harnesses LLMs for meme creation and thoroughly assesses their capability to produce humorous memes.

\begin{figure*}[t]
  \centering
  \includegraphics[width=0.95\textwidth]{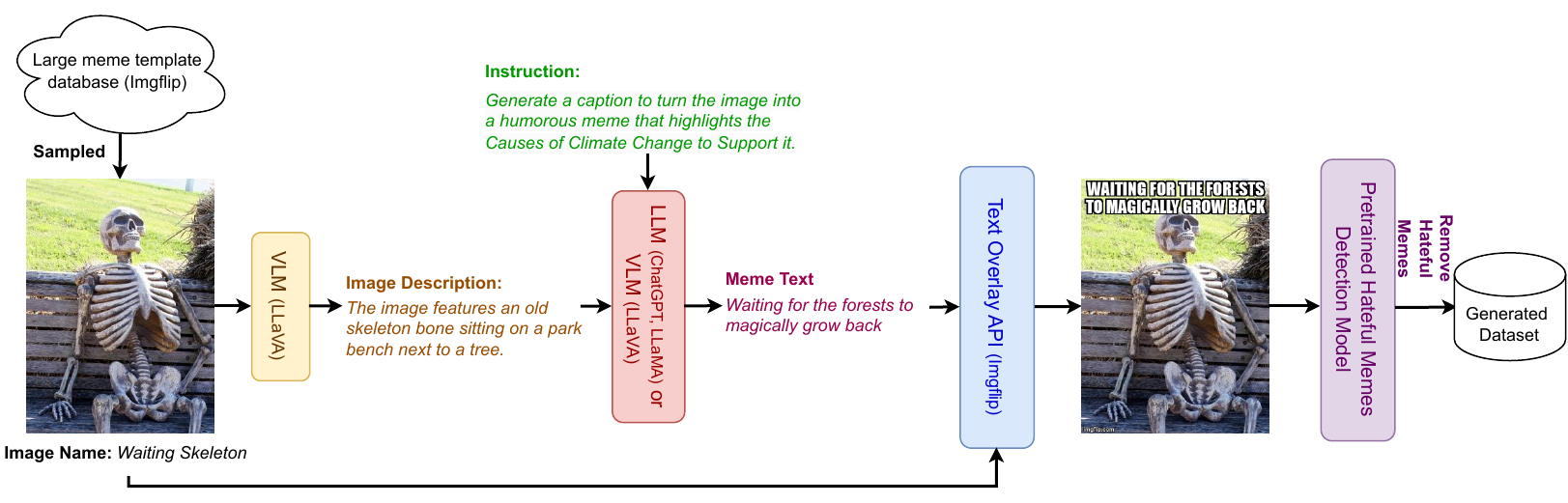}
  \caption{Overview of \textbf{MemeCraft} framework.}
  \label{fig:meme_gen_framework}
\end{figure*}

Recent advancements in the vision-language domain have led to the emergence of large vision-language models (VLMs) known for their robust generative abilities~\cite{ye2023mplug, dai2023instructblip, liu2023visual, zhu2023minigpt,li-blip-2-2023, openai-gpt4-2023b}. These nascent VLMs set themselves apart from the prevailing smaller versions by virtue of their extensive model size and superior performance in multimodal understanding tasks. Recent research indicates that VLMs demonstrate noteworthy zero-shot capabilities across various multimodal tasks~\cite{zheng2023judging}, highlighting their potential applicability in diverse downstream applications such as image captioning. Examples of these advancements include ChatCaptioner~\cite{zhu-chatgpt-blip-2-2023}, which integrates ChatGPT with BLIP-2 for image captioning, and LLaVA-Med~\cite{li-llava-med-2023}, which utilizes GPT-4 for queries related to biomedical images. Drawing inspiration from these existing studies, we leverage LLaVA~\cite{liu2023visual} visual comprehension and text generation prowess for meme creation.

\section{Methodology}

Figure~\ref{fig:meme_gen_framework} presents an overview of our \textsf{MemeCraft} framework. As most viral memes are generated based on popular meme template images, we first sample meme template images from online repositories like Imgflip. The sampled images are then processed using a VLM to derive their textual descriptions (i.e., captions). Following this, we input the name of the meme template and generated descriptions into either an LLM or VLM to produce humorous meme texts that align with specific social causes, such as "\textit{climate action}" and "\textit{gender equality}. To create the final memes, the generated meme texts are superimposed on the sampled meme template images using a text overlay API. Finally, to ensure that our proposed framework does not generate hateful memes, we use a pre-trained model dedicated to detecting hateful memes to filter the generated memes. Only memes deemed non-hateful are kept in our generated dataset. The details of each 
\textsf{MemeCraft} component are discussed in the subsequent subsections.


\subsection{Meme Template Image Sampling}
For effective application of our model, it is essential to have access to meme templates. These templates are pre-existing images commonly used for meme generation, often accompanied by a unique name and a specific format for text addition, as provided by meme creation websites.

We built our collection of meme templates using the \textit{Meme Generator Dataset}\footnote{\url{https://www.kaggle.com/datasets/electron0zero/memegenerator-dataset}}, identifying 1914 distinct templates. It's worth noting that the blank image URLs in this dataset were no longer functional. Therefore, we retrieved 1500 meme template IDs from the Imgflip API, using the image names from the dataset. These IDs are crucial for sourcing the blank image URLs and enabling text overlay through the Imgflip API.



\subsection{Image Description Generation}
For the task of image description, we utilize the LLaVA-7B, the 7 billion parameter version of LLaVA~\cite{liu2023visual}, a state of the art VLM. LLaVA is a comprehensively trained multimodal model that encompasses both visual and language comprehension abilities, and existing research have showed that LLaVA is able to perform well in image captioning task \cite{zhai2023hallE-switch}. Given its adeptness in generating detailed image captions, we leverage the VLM by issuing a zero-shot prompt to obtain high-quality image descriptions for each meme template image. Figure~\ref{fig:meme_gen_framework} shows an illustrative example of a meme tempalte image and its corresponding generated image description.


\subsection{Meme Text Generation}
The goal of this task is to produce meme text that, when combined with a meme template image, results in a multimodal meme that communicates a campaign message. This message is guided by three user-defined parameters:

\begin{itemize}
    \item \textbf{Social Cause}: This denotes the social issue or movement the meme addresses. While our experiments focused on two causes, namely ``\textit{climate action}'' and ``\textit{gender equality}'', it is noteworthy that \textsf{MemeCraft} can generate memes for a broader range of social topics.
    \item \textbf{Stance}: This determines whether the meme will \textit{support} or \textit{deny} the chosen social cause.
    \item \textbf{Persuasion Technique}: This specifies the method employed to sway the audience towards supporting or opposing the social cause. As an example, a strategy for championing climate action might highlight the consequences of climate change. Although several persuasion methods are versatile and fit various causes, some causes might necessitate specialized techniques. The specific techniques used for ``\textit{climate action}'' and ``\textit{gender equality}'' are elaborated in the next section.
\end{itemize}

To generate meme text that conveys the intended social campaign message, we employ prompts based on the three user-defined parameters and pair them with randomly selected meme template images. An exemplary prompt supporting ``\textit{climate action}'' using the \textit{causes of climate change} as a persuasion technique is demonstrated in Table~\ref{tab:prompt_design}. We enhance this prompt with the prefix ``\textit{Let's think step by step}.''~\cite{zhang2022chain}. This prefix encourages LLMs to establish a stronger connection between the image and meme text, ensuring the generated text closely aligns with the visual content.

\definecolor{question_color}{HTML}{1B9E77}
\definecolor{label_color}{HTML}{D95F02}
\definecolor{explanation_provided_color}{HTML}{7570B3}
\definecolor{explanation_generated_color}{HTML}{FF55A3}
\definecolor{tweet_color}{HTML}{FFC300}

\begin{table}[t!]
\centering
\small
\caption{An illustration of a prompt utilizing $N$ training examples is presented, consisting of the instruction (in \textcolor{question_color}{green}), the input (in \textcolor{tweet_color}{yellow}), and the output provided (in \textcolor{explanation_provided_color}{purple}) that are utilized as input to LLMs.}
\label{tab:prompt_design}
\begin{tabular}{p{0.90\linewidth}}
\toprule
\# Demonstration samples $1 ... N$\\
\textbf{Instruction}: \textcolor{question_color}{Generate a caption to turn the image into a humorous meme that highlights the Causes of Climate Change to Support it.}\\

\textbf{Input}: \textcolor{tweet_color}{Image "willywonka" describing "A man wearing a purple and gold suit, a top hat, and..."}\\

\textbf{Output}: \textcolor{explanation_provided_color}{[Let's think step-by-step.] The "willywonka" image is frequently employed sarcastically, often with rhetorical questions. Let's use it for... \textbf{Caption at top: "You think global warming is fake?" and Caption at bottom: "Please tell me how you get the "FACTS" from politicians and oil companies"}}\\
\#\#\#\\
\# Actual image in prompt\\

\textbf{Instruction}: \textcolor{question_color}{Generate a caption to turn the image into a humorous meme that highlights the Causes of Climate Change to Support it.}\\

\textbf{Input}: \textcolor{tweet_color}{Image "Waiting Skeleton" describing "The image features an old skeleton bone sitting on a park bench..."}\\

\textbf{Output}: \textcolor{explanation_provided_color}{[Let's think step-by-step.]}
\textcolor{explanation_generated_color}{The "Waiting Skeleton" image is often used to depict patience or waiting with a touch of irony. Let's use this image for... \textbf{Caption at top: "Waiting for the forests to magically grow back"}}\\
\bottomrule
\end{tabular}
\end{table}


In our study, we utilize three models as variations of \textsf{MemeCraft} to generate humorous meme text based on our specifically designed prompts: ChatGPT-3.5\footnote{\url{https://api.openai.com/v1/chat/completions}}, LLaMA-2-13B~\cite{touvron2023llama}, and LLaVA-7B~\cite{liu2023visual}.

For the LLMs, namely ChatGPT and LLaMA, we adopt a few-shots prompting technique, illustrated in Table~\ref{tab:prompt_design}. Through various tests, we evaluated the effectiveness of providing different numbers of demonstration examples, denoted as $N$, varying from 2 to 8. Our observations revealed that when $N$ is minimal, such as 2, LLaMA faces challenges in comprehending the task. On the other hand, when $N$ is increased to values like 6 or 8, LLaMA tends to mimic the provided examples. As a result, we empirically settled on presenting 4 demonstration examples. 

In contrast, LLaVA, a VLM, does not support the few-shots prompting technique. Hence, we employed a zero-shot prompting method for this model. While the structure of the prompting format aligns with the template presented in Table~\ref{tab:prompt_design}—comprising \textit{Instruction}, \textit{Input}, and \textit{Output} segments—it does not incorporate any demonstration samples. Additionally, the prefix ``\textit{Let's think step-by-step}'' is excluded from the output, prompting the VLM to generate the meme text directly.

\subsection{Hateful Memes Detection}

Recent research indicates a concerning rise in the production and distribution of hateful memes, which not only target, but also risk inciting violence against protected groups~\cite{kiela2020hateful,mathias2021fhmfg}. In light of this, we have enhanced \textsf{MemeCraft} with a safety mechanism to mitigate the risk of generating such content. This mechanism involves the critical evaluation of generated memes for potential hateful content.

Central to our approach is the employment of the MultiModal BiTransformers (MMBT-Grid) model, a dedicated hateful memes detection technique introduced in \cite{kiela2019supervised}. MMBT-Grid adopts a multimodal supervised bitransformer architecture, which utilizes distinct unimodal components. These components have been trained to associate multimodal image embeddings with textual tokens.

Our choice of MMBT-Grid is supported by its thorough pre-training on the Hateful Meme Challenge (HMC) Dataset, a prominent benchmark for identifying hateful content \cite{kiela2020hateful}. MMBT-Grid achieves 66.85\% accuracy on this dataset \cite{van2023detecting}. To balance the generation of diverse memes while prioritizing safety, we established a high confidence threshold of 0.9 to classify content as hateful. Any content exceeding this threshold is excluded. Memes that are validated through this mechanism are added to our final dataset.



\section{Dataset Generation}
In this section, we present the process of utilizing \textsf{MemeCraft} to generate a dataset that focuses on two social causes: ``\textit{climate action}'' and ``\textit{gender equality}''. The generated dataset will be used in our large-scale human evaluation in the subsequent section. 

\begin{table*}
 \small
  \caption{Distribution of memes generated for human evaluation.}
  \begin{tabular}{ccccccccc}
    \toprule
    Social Cause &Model&Causes&Consequences&Solutions&Evidence of Absence&Benefits&Rationale&Total\\
    \midrule
    \multirow{3}{*}{Climate Action} &  \textit{MemeCraft}-ChatGPT & 100& 100& 100& 100& 100&  0& 500\\
     & \textit{MemeCraft}-LLaMA& 100& 100& 100& 100& 100& 0& 500\\
    & \textit{MemeCraft}-LLaVA & 100& 100& 100& 100& 100& 0& 500 \\
    \midrule
    \multirow{3}{*}{Gender Equality} &  \textit{MemeCraft}-ChatGPT & 100& 100& 100& 100&0  & 100& 500  \\
     & \textit{MemeCraft}-LLaMA & 100& 100& 100& 100&0  & 100& 500  \\
    & \textit{MemeCraft}-LLaVA &100& 100& 100& 100&0  & 100& 500 \\
     \midrule
    - & Dank Learning &0 &0 &0 &0&0 &0 & 100  \\
     \midrule
   - & Online Random &0 &0 &0 &0&0 &0 & 100 \\
    
  \bottomrule
\end{tabular}
\label{tab:num_of_memes}
\end{table*}

For each social cause, we first design prompts with persuasion techniques to generate the memes. These memes aim to influence people to support or oppose the given cause. Specifically, we use the following techniques for persuasion:

\textbf{Causes}: This technique emphasizes factors that can contribute to or intensify the social causes. This approach is used in supporting ``\textit{climate action}'' and ``\textit{gender equality}''. An illustrative example involves a meme that accuses individuals of failing to switch off lights every time they exit a room in the context of ``\textit{climate action}''.

\textbf{Consequences}: This technique emphasizes the potential negative consequences that may arise from underlying social causes. This approach is used in supporting ``\textit{climate action}'' and ``\textit{gender equality}''. An illustrative example includes a meme that emphasizes how global warming affects everyone, not just polar bears, within the context of ``\textit{climate action}''.

\textbf{Solutions}: This technique offers solutions and encourages individuals to take action in addressing the social causes. This approach is used in supporting ``\textit{climate action}'' and ``\textit{gender equality}''. An illustrative example involves a meme that highlights the idea that single-use plastic can be environmentally friendly if it's reused, within the context of ``\textit{climate action}''.

\textbf{Evidence of Absence}: This technique presents evidence suggesting that the social causes either do not exist or are not significant problems.  This approach is used in denying ``\textit{climate action}'' and ``\textit{gender equality}''. An illustrative example includes a meme that emphasizes the recent increase in the number of polar bears within the context of ``\textit{climate action}''.

\textbf{Benefits}: This technique highlights the idea that the social causes could have positive effects for us. This approach is used in denying ``\textit{climate action}''. An illustrative example includes a meme that emphasizes how farmers can now cultivate plants year-round due to global warming.

\textbf{Rationale}: This technique emphasizes that the existence of a social causes is supported by valid reasons. This approach is used in denying ``\textit{gender equality}''. An example includes a meme that underscores the notion that because women and men have distinct physical differences, it is natural to treat them differently.


After establishing the persuasion technique, we generate memes focused on ``\textit{climate action}'' and ``\textit{gender equality}'' using the three variations of \textsf{MemeCraft}: ChatGPT, LLaMA, and LLaVA. Each \textsf{MemeCraft} variation produces 500 memes for each social cause. The distribution of the generated memes for our human evaluation is displayed in Table \ref{tab:num_of_memes}. In total, \textsf{MemeCraft} was used to produce 3,000 memes. For comparison, we also created 100 memes with the state-of-the-art Dank Learning meme generator~\cite{peirson2018dank}, and randomly sampled 100 online memes from ImgFlip.

\section{Experiments}
In this section, we describe the comprehensive human evaluation conducted with the generated dataset. Specifically, we designed experiments to address the following research questions:

\begin{itemize}
    \item \textbf{R1}: Do the generated memes resemble publicly available online memes? [\textbf{\textit{Authencity}}]
    \item \textbf{R2}: Are the generated memes humorous? [\textbf{\textit{Hilarity}}]
    \item \textbf{R3}: Do the generated memes communicate the intended message? (e.g., support climate action) [\textbf{\textit{Message Conveyance}}]
    \item \textbf{R4}: Are the generated memes persuasive? [\textbf{\textit{Persuasiveness}}]
    \item \textbf{R5}: Is the safety mechanism effectiveness in reducing hateful meme generation? [\textbf{\textit{Hatefulness}}]
\end{itemize}

\subsection{Experimental Settings}

\textbf{Human Evaluators.} We recruited a group of 12 undergraduate students, selected for their familiarity with online meme culture, to serve as evaluators in our study. Each meme was assessed by at least two evaluator, and the average rating from these evaluations was recorded to represent the consensus on each meme.


\textbf{Evaluation Metrics.} To address the research questions outlined in our study, we developed five evaluation metrics to assess the various aspects of the generated memes comprehensively. 

\textit{Authenticity}: This metric gauges the resemblance of generated memes to those typically encountered online. Evaluators determinte whether a meme resembled those found on the internet, with responses categorized as ``\textit{yes}'' or ``\textit{no}''. The \textit{Authenticity} score is computed as the percentage of memes rated as ``\textit{yes}'',

\textit{Hilarity}: This metric evaluates the humor of the generated meme. Evaluators rate the humor of meme using a 5-point scale, with 1 indicating ``\textit{not humorous}'' and 5 indicating ``\textit{humorous}''.

\textit{Message Conveyance}: This metric evaluates the capacity of generated memes to adeptly convey the intended message, particularly in terms of supporting or denying specific social causes. Evaluators categorize each meme as either ``\textit{support}'', ``\textit{deny}'', or exhibiting no clear alignment (``\textit{NA}'') with the specified social cause. The \textit{Message Conveyance} score is compute by the percentage of memes evaluated to convey the specified social cause and stance. 


\textit{Persuasiveness}: This metric measures the persuasive quality of the generated memes. Evaluators rate the persuasiveness of memes on a scale from 1 to 5, where 1 signified ``\textit{not persuasive}'' and 5 signified ``\textit{persuasive}''.

\textit{Hatefulness}: This metric assesses the effectiveness of a safety mechanism in mitigating the generation of hateful memes. The safety mechanism classifies generated memes as either \textit{hateful} or \textit{non-hateful} and filters out the hateful ones. Evaluators are tasked to identify the presence of hateful content within the remaining memes. We report two \textit{hatefulness} scores: one represents the percentage of hateful memes identified by machine (i.e., safety mechanism), and the other denotes the percentage of hateful memes identified by the evaluators.

\begin{table}[t]
\centering
\small
  \caption{Average authenticity scores of various models.}
  \label{tab:similarity}
  \begin{tabular}{ccc}
    \hline
    \textbf{Social Cause} &\textbf{Model}&\textbf{Authenticity Score}\\
    \hline
   \multirow{3}{*}{Climate Action} &  \textsf{MemeCraft}-ChatGPT & \textbf{0.48}\\
     & \textsf{MemeCraft}-LLaMA& 0.43 \\
     & \textsf{MemeCraft}-LLaVA & 0.36 \\
    \hline
   \multirow{3}{*}{Gender Equality} &  \textsf{MemeCraft}-ChatGPT & \textbf{0.53}  \\
      & \textsf{MemeCraft}-LLaMA & 0.45  \\
     & \textsf{MemeCraft}-LLaVA & 0.34  \\
     \hline
     - & Dank Learning & 0.23  \\
     \hline
    - & Online Random & \textbf{0.59} \\
  \hline
\end{tabular}
\end{table}

\subsection{Authenticity}

Table~\ref{tab:similarity} shows the average authenticity scores of the various meme generators. Specifically, we report the three variations of our \textsf{MemeCraft} model and the two baseline methods. These scores are indicative of each generator's ability to create memes that are perceived as genuine when compared to typical online content.

Interestingly, we observed that even authentic online memes (i.e., \textsf{Online Random}) achieved an authenticity score of only 0.59. However, it was noted that \textsf{MemeCraft}-ChatGPT attained authenticity scores closer to those of authentic online memes, with averages of 0.48 for ``\textit{climate action}'' and 0.53 for ``\textit{gender equality}''. In contrast, due to their smaller model sizes, \textsf{MemeCraft}-LLaMA and \textsf{MemeCraft}-LLaVA achieved lower scores of approximately 0.45 and 0.35, respectively. Notably, all three \textsf{MemeCraft} model variations significantly outperformed the \textsf{Dank Learning} baseline.

The authenticity scores for the ``\textit{gender equality}'' theme are observed to be slightly higher than those for the ``\textit{climate action}''. The richness of content available for ``\textit{gender equality}'' themes offers a wealth of recognizable symbols, narratives, and humor that models can draw from to craft their memes. This richness allows for a diverse expression that aligns closely with the authentic online memes. Furthermore, the complexity of the ``\textit{climate action}'' topic may not translate as effectively into the meme format, which often relies on simplicity and immediacy for impact. The ``\textit{gender equality}'' memes might encapsulate their message more straightforwardly, avoiding the intricate scientific explanations that can dilute the authenticity of ``\textit{climate action}'' memes.

\subsection{Hilarity}

\begin{figure}[t]
  \centering
  \includegraphics[width=0.415\textwidth]{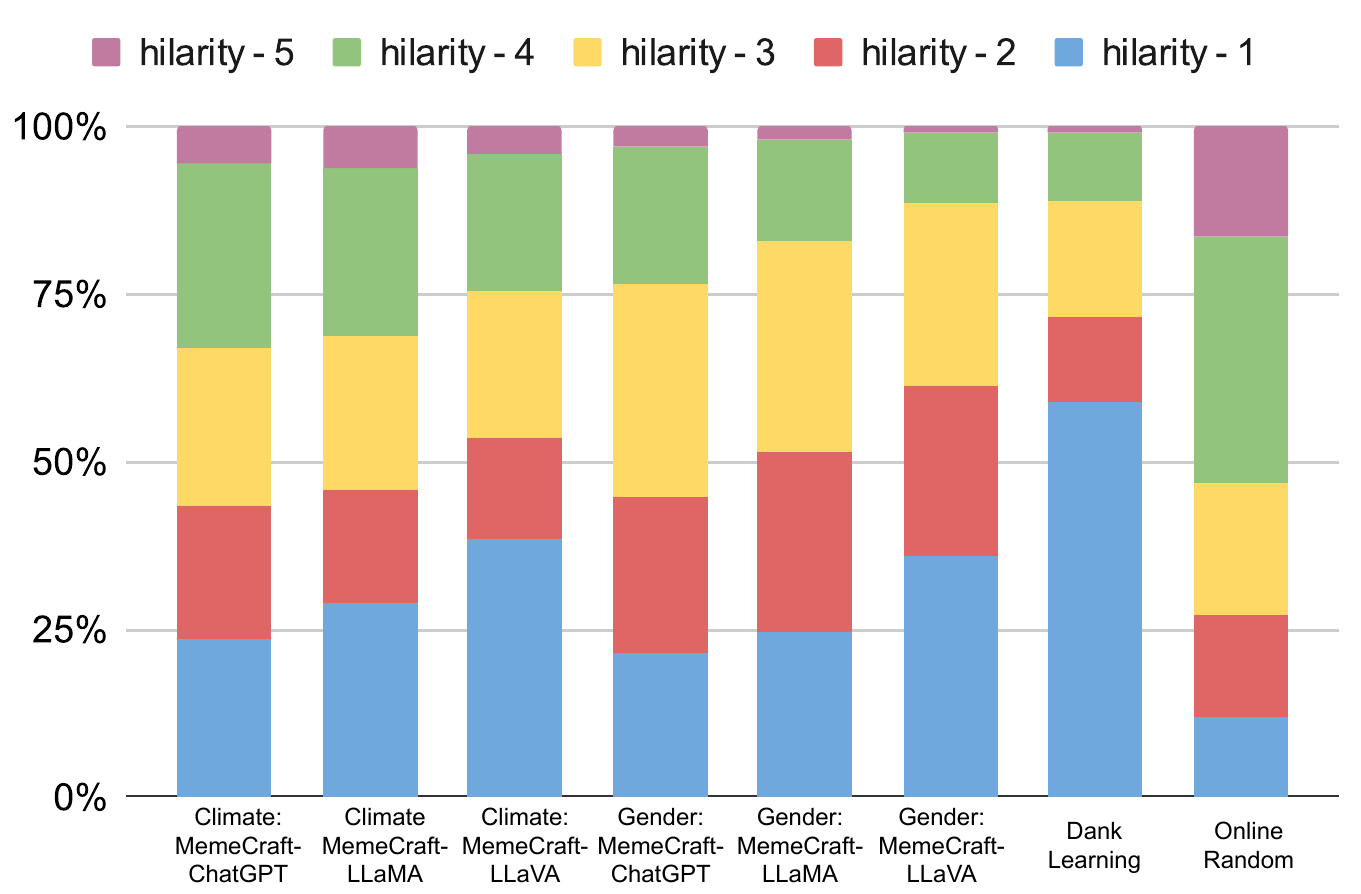}
  \caption{Distribution of hilarity scores across models and social causes}
  \label{fig:hilarity_overall}
\end{figure}

Figure \ref{fig:hilarity_overall} presents the distribution of hilarity scores across different social causes and meme generation models. The analysis highlights that \textsf{MemeCraft}-ChatGPT is one standout model, achieving a median hilarity score of 3 in both social causes examined. This score suggests that over half of the memes generated by \textsf{MemeCraft}-ChatGPT were considered to have a moderate level of humor. We also noted that the LLM-based models, \textsf{MemeCraft}-ChatGPT and \textsf{MemeCraft}-LLaMA, perform better than the VLM-base model, \textsf{MemeCraft}-LLaVA. The observed performance disparity can be attributed to the inherent limitations of the LLaVA model, which exclusively relies on zero-shot learning capabilities. Unlike other models that can learn and improve from demonstration examples, LLaVA operates without such examples in its prompts, leading to comparatively suboptimal performance. Comparatively, the baseline model \textsf{Dank Learning} yielded a median hilarity score of only 1, indicating a general lack of humor its generated memes. Nevertheless, machine-generated memes have lower hilarity scores compared to authentic online memes. The authentic online memes (i.e., \textsf{Online Random}) have a median hilarity score of 4. This finding underscores that there remains room for improvement for machines to generate humorous content.

\subsection{Message Conveyance}

\begin{table*}
\small
  \caption{Message conveyance scores of memes created by supportive and denied persuasion techniques.}
  \label{tab:conveyance}
  \begin{tabular}{ccccccccc}
    \hline
    &&&\multicolumn{3}{c}{ \textbf{Supportive Persuasion Techniques}} &\multicolumn{3}{c}{ \textbf{Denied Persuasion Techniques}} \\
     \cmidrule(l{2pt}r{2pt}){4-6}\cmidrule(l{2pt}r{2pt}){7-9}
    \textbf{Stance Evaluated}&\textbf{Social Cause} &\textbf{Model}& Causes&Consequences&Solutions&Evidence of Absence&Benefits&Rationale\\
    \hline
    \multirow{6}{*}{Support} & \multirow{3}{*}{Climate Action} &  MemeCraft-ChatGPT &  0.71& 0.65& 0.8& 0.3& 0.43& -\\
    & & MemeCraft-LLaMA&  0.71& 0.47& 0.45& 0.17& 0.45& - \\
    & & MemeCraft-LLaVA  & 0.51& 0.55& 0.42& 0.24& 0.38&-  \\
    \cline{2-9}
    & \multirow{3}{*}{Gender Equality} &  MemeCraft-ChatGPT &  0.76& 0.63& 0.67& 0.41& -& 0.51  \\
     & & MemeCraft-LLaMA &  0.69& 0.54& 0.6& 0.34& -& 0.43  \\
    & & MemeCraft-LLaVA &   0.42& 0.38& 0.48& 0.29& -& 0.34  \\
   \hline
    \multirow{6}{*}{Deny} & \multirow{3}{*}{Climate Action} &  MemeCraft-ChatGPT&  0.1& 0.17& 0.05& 0.5& 0.4& -\\
     && MemeCraft-LLaMA&  0.1& 0.07& 0.12& 0.76& 0.39& -\\
    && MemeCraft-LLaVA  &  0.15& 0.1& 0.12& 0.43& 0.28& -\\
     \cline{2-9}
    &\multirow{3}{*}{Gender Equality} &  MemeCraft-ChatGPT &  0.08& 0.1& 0.06& 0.49& -& 0.33  \\
    & & MemeCraft-LLaMA &  0.11& 0.1& 0.04& 0.28& -& 0.39  \\
   & & MemeCraft-LLaVA &  0.07& 0.04& 0.08& 0.2& -& 0.17  \\
  \hline
\end{tabular}
\end{table*}

Table \ref{tab:conveyance} presents the message conveyance scores of the three \textsf{MemeCraft} variations when employing different persuasion techniques to either support or oppose the social causes. 

In our examination of memes evaluated as supporting the social causes, we observed notable differences among various persuasion strategies. The technique we have termed ``\textit{Causes}'' emerged as the most effective, yielding approximately 0.7 supportive memes when utilized by \textsf{MemeCraft}-ChatGPT and \textsf{MemeCraft}-LLaMA, i..e, among the memes generated using ``\textit{Causes}'' technique, about 70\% of these memes are evaluated as supporting the social cause. Additionally, other supportive strategies registered efficacy scores exceeding 0.6 with \textsf{MemeCraft}-ChatGPT. In contrast, persuasion strategies specifically designed to deny social causes produced approximately 30\% of memes that still advocated for the cause.


In our analysis of memes evaluated as denying social causes, we discovered that memes utilizing supportive persuasion techniques rarely adopted an opposing viewpoint, with only approximately 10\% doing so. Conversely, the efficacy of various denied persuasion techniques generally yielded a low conversion to contrarian memes. A significant outlier was observed with the ``\textit{Evidence of Absence}'' technique in the context of the ``\textit{climate action}'' social cause when applied in the \textsf{MemeCraft}-LLaMA model. Here, 76\% of the generated memes opposed the cause. This anomaly was attributed to the \textsf{MemeCraft}-LLaMA model's replication of patterns from provided demonstration examples during the meme generation process.

\subsection{Persuasiveness}
\begin{figure*}[t]
  \centering
  \includegraphics[width=\textwidth]{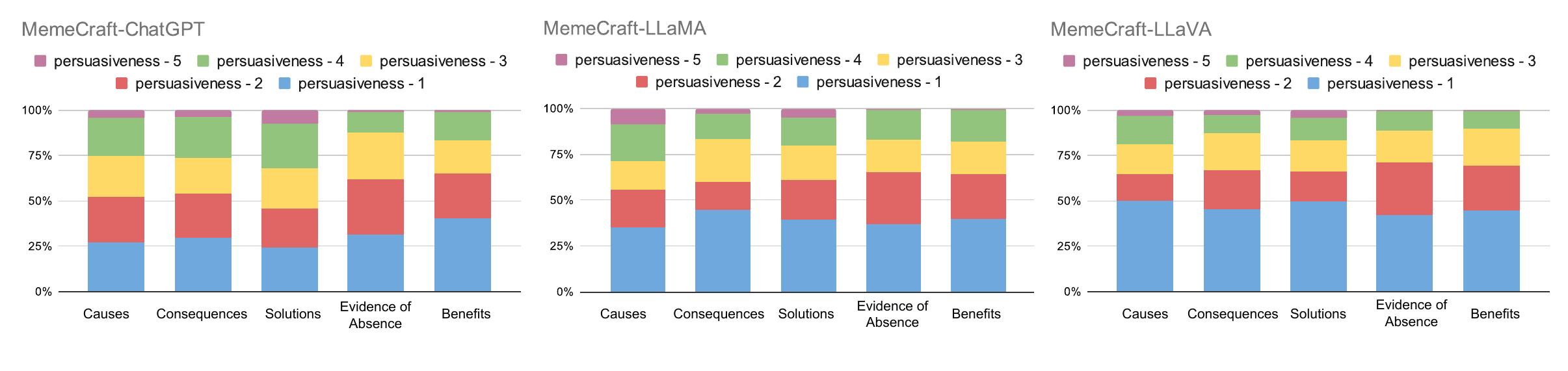}
  \caption{Persuasiveness scores distribution of various models generating climate action memes.}
  \label{fig:persuasiveness_climate}
\end{figure*}

\begin{figure*}[t]
  \centering
  \includegraphics[width=\textwidth]{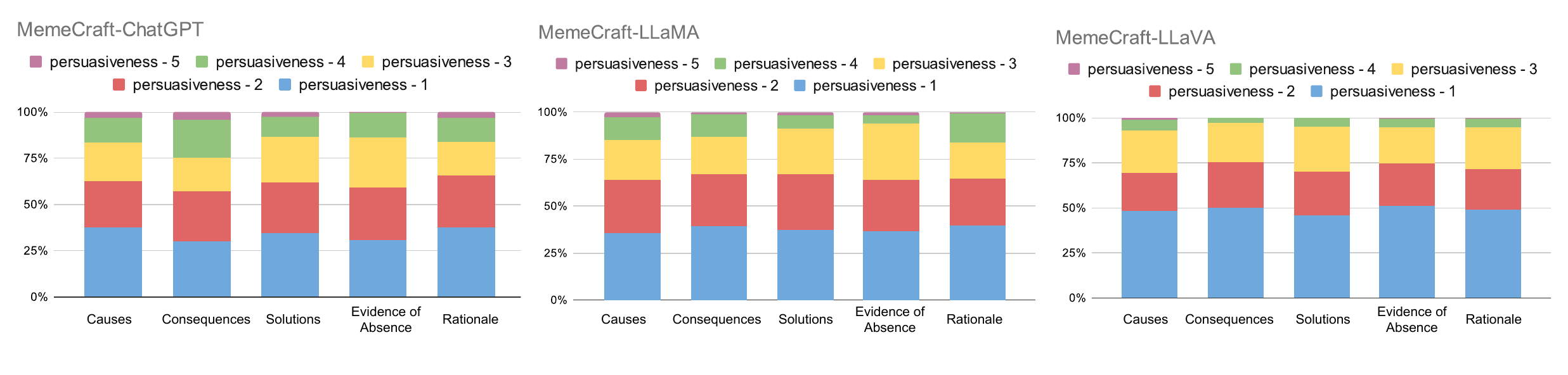}
  \caption{Persuasiveness scores distribution of various models generating gender equality memes.}
  \label{fig:persuasiveness_gender}
\end{figure*}

Figure~\ref{fig:persuasiveness_climate} and \ref{fig:persuasiveness_gender} present the persuasiveness score distributions of various models generating ``\textit{climate action}'' and ``\textit{gender equality}'' memes, respectively. Overall, we observed that \textsf{MemeCraft}-ChatGPT outperformed other \textsf{MemeCraft} models in both social causes, with around 40\% of memes receiving a rating of 3 out of 5 persuasiveness score. It is noteworthy that different persuasion techniques could exhibit varying levels of performance. In general, within the social cause of ``\textit{climate action}'', persuasion techniques emphasizing support tended to perform better. In the context of the ``\textit{Solutions}'' technique employed within \textsf{MemeCraft}-ChatGPT, it notably attained a persuasive score of 3 for more than 50\% of the evaluated memes. Conversely, in the social cause of ``\textit{gender equality}'', the disparity in persuasiveness scores between support and deny persuasion techniques was less pronounced. 

In summary, the memes we generated demonstrate moderate persuasive, and the persuasiveness scores are notably influenced by the social causes and persuasion techniques used. These findings suggest that tailoring persuasion techniques to specific social causes is crucial for enhancing the persuasive impact of the memes.


\subsection{Hatefulness}

\begin{table}[t]
\small
  \caption{Hatefulness scores of generated memes identified by the safety mechanism (\textit{Machine}) and post-filtered memes assess by the evaluators (\textit{Human}).}
  \label{tab:hatefulness_model}
  \begin{tabular}{cccc}
    \toprule
    Social Cause &Model&Machine&Human\\
    \midrule
    \multirow{3}{*}{Climate Action} &  MemeCraft-ChatGPT & 0.06&0.01\\
     & MemeCraft-LLaMA & 0.03&0 \\
    & MemeCraft-LLaVA &  0.03&0.01 \\
    \midrule
    \multirow{3}{*}{Gender Equality} &  MemeCraft-ChatGPT  & 0.55&0.02  \\
     & MemeCraft-LLaMA & 0.41&0.04 \\
    & MemeCraft-LLaVA  & 0.27&0.01 \\
    \midrule
        - & Dank Learning & - & 0.05  \\
     \midrule
    - & Online Random &  - & 0.05 \\
  \bottomrule
\end{tabular}
\end{table}

Table~\ref{tab:hatefulness_model} presents two sets of hatefulness scores: (i) hateful memes identified by \textsf{MemeCraft}'s safety mechanism ( \textit{machine}), and (ii) the hateful memes identified by the evaluators in the set post-filtered by the safety mechanism. In contentious social issues like "gender equality," our safety mechanism notably detects a markedly higher proportion of memes as hateful, highlighting the risk for LLMs to propagate hate speech~\cite{kumar2024ethics}. This underscores the need for caution when creating memes in such social causes.

Following the removal of memes identified as hateful by our models, we engaged human annotators to assess the remaining memes for hatefulness. After applying our safety mechanism, we found that the percentage of hateful memes in both social causes consistently remained below 0.05, which is lower than that of two baselines. While ``\textit{gender equality}'' social cause had a slightly higher proportion of hateful memes, the percentages still stayed within acceptable limits. This serves as empirical evidence affirming the efficacy of our framework for filtering out hateful memes.

\begin{table*}
  \centering
  \small
  \begin{tabular}
  {p{1.7cm}|p{2.8cm}|p{3.5cm}|p{4.0cm}|p{3.9cm}}
  \toprule 
      \multirow{2}{*}{\textbf{Memes}}
    &
    \begin{minipage}[!b]{0.1\columnwidth}
  \centering
{\includegraphics[width=2.9cm, height=3cm]{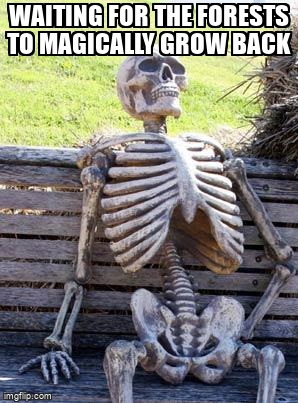}}
 \end{minipage}  
 &
    \begin{minipage}[!b]{0.1\columnwidth}
  \centering
 {\includegraphics[width=3.5cm, height=3cm]{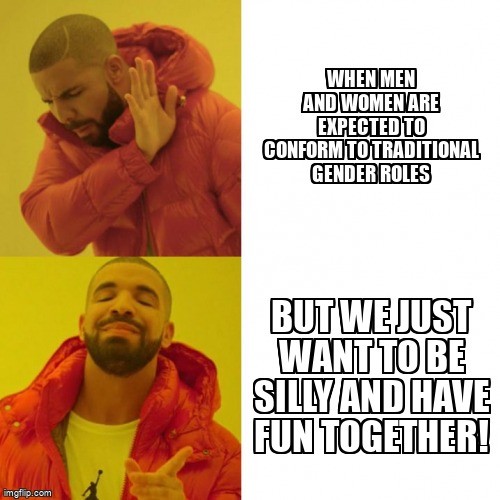}}
 \end{minipage} 
  &
    \begin{minipage}[!b]{0.1\columnwidth}
  \centering
{\includegraphics[width=4.0cm, height=3cm]{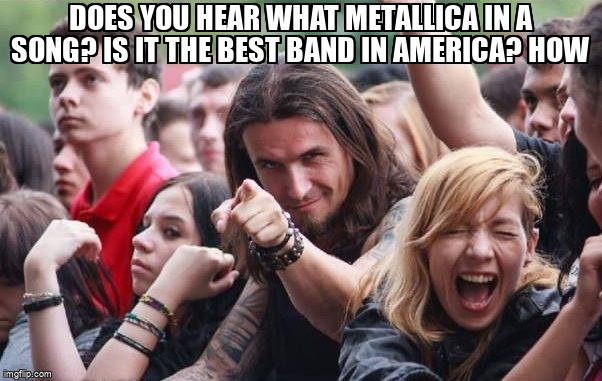}}
 \end{minipage} 
 &
    \begin{minipage}[!b]{0.1\columnwidth}
  \centering
  {\includegraphics[width=3.9cm, height=3cm]{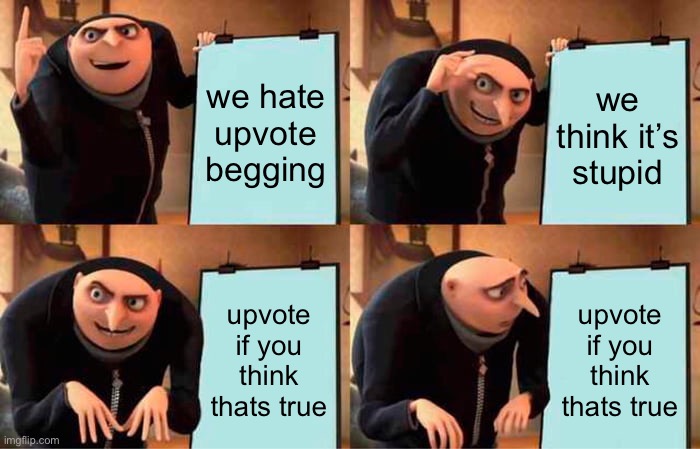}}
 \end{minipage}
 \\
   & (1) &  (2) & (3)  &  (4)\\
    \hline

   \textbf{Meme Texts} & \textit{Waiting for the forests to magically grow back} & \textit{When men and women are expected to conform to traditional gender roles - But we just want to be silly and have fun together!} & \textit{Does you hear what metallica in a song? It the best band in America? How} & \textit{we hate upvote begging - we think it's stupid - upvote if you think thats true - upvote if you think thats true} \\  \hline
   
    \multirow{2}{*}{\textbf{Evaluations}}& \{A: \textcolor{blue}{Yes}, M-H: \textcolor{blue}{5}, M: \textcolor{blue}{Support}, P: \textcolor{blue}{5}, H: \textcolor{blue}{No}\} & \{A: \textcolor{blue}{Yes}, M-H: \textcolor{blue}{5}, M: \textcolor{blue}{Support}, P: 
    \textcolor{blue}{5}, H: \textcolor{blue}{No}\} & \{A: \textcolor{blue}{Yes}, M-H: \textcolor{blue}{4}, H: \textcolor{blue}{No}\} & \{A: \textcolor{blue}{Yes}, M-H: \textcolor{blue}{5}, H: \textcolor{blue}{No}\}\\  \cline{2-5}
     & \{A: \textcolor{blue}{Yes}, M-H: \textcolor{blue}{4}, M: \textcolor{blue}{Support}, P: \textcolor{blue}{4}, H: \textcolor{blue}{No}\} & \{A: \textcolor{blue}{Yes}, M-H: \textcolor{blue}{4}, M: \textcolor{blue}{Support}, P: \textcolor{blue}{3}, H: \textcolor{blue}{No}\} & \{A: \textcolor{red}{No}, M-H: \textcolor{blue}{3}, H: \textcolor{blue}{No}\} & \{A: \textcolor{blue}{Yes}, M-H: \textcolor{blue}{5}, H: \textcolor{blue}{No}\} \\  \hline

       \textbf{Social Causes and Stances} & Support Climate Action & Support Gender Equality & - & -  \\  \hline
  \textbf{Models} & MemeCraft-LLaMA & MemeCraft-ChatGPT & Dank Learning & Random Online  \\  
  


  \toprule 

      \multirow{2}{*}{\textbf{Memes}}
    &
    \begin{minipage}[!b]{0.1\columnwidth}
  \centering
{\includegraphics[width=2.9cm, height=3cm]{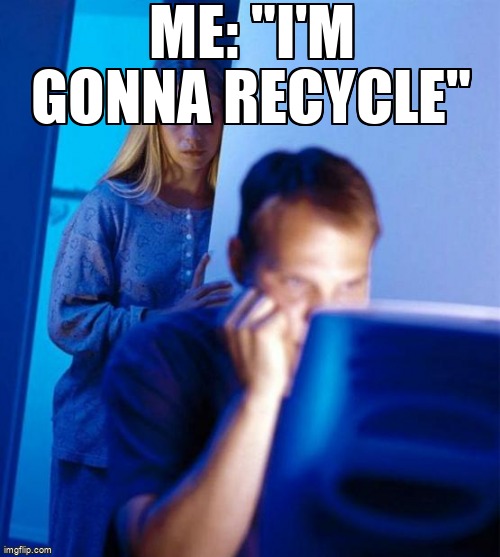}}
 \end{minipage}  
 &
    \begin{minipage}[!b]{0.1\columnwidth}
  \centering
 {\includegraphics[width=3.4cm, height=3cm]{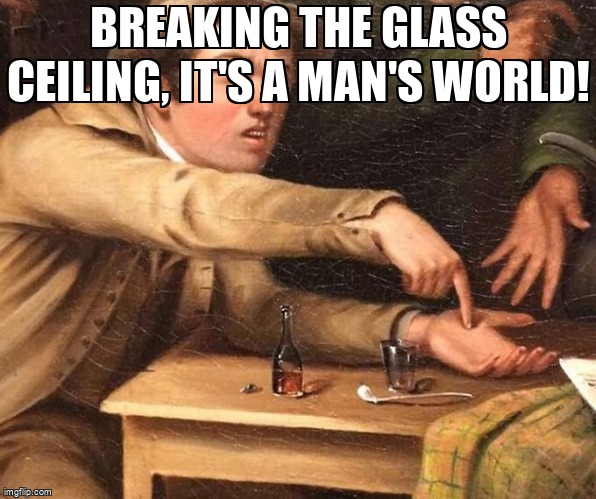}}
 \end{minipage} 
  &
    \begin{minipage}[!b]{0.1\columnwidth}
  \centering
{\includegraphics[width=4.0cm, height=3cm]{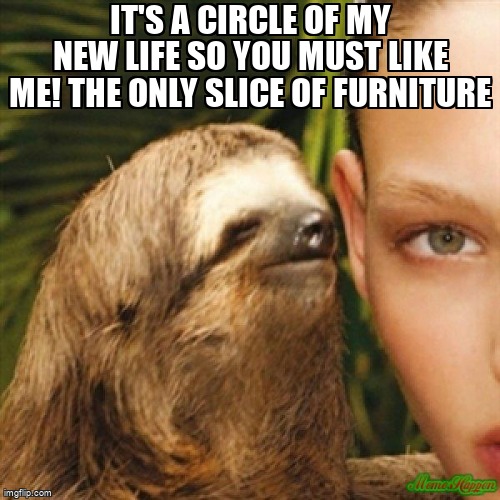}}
 \end{minipage} 
 &
    \begin{minipage}[!b]{0.1\columnwidth}
  \centering
  {\includegraphics[width=3.9cm, height=3cm]{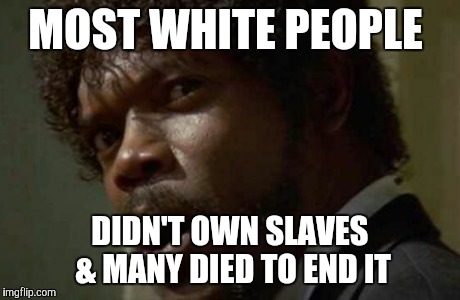}}
 \end{minipage}
 \\
   & (5) &  (6) & (7)  &  (8)\\
    \hline 
    

       \textbf{Meme Texts} & \textit{Me: "I'm gonna recycle"} & \textit{Breaking the glass ceiling, it's a man's world!} & \textit{It's a circle of my new life so you must like me! The only slice of furniture} & \textit{Most white people - Didn't own slaves \& Many died to end it} \\  \hline
       
     \multirow{2}{*}{\textbf{Evaluations}} & \{A: \textcolor{red}{No}, M-H: \textcolor{red}{2}, M: \textcolor{blue}{Support}, P: \textcolor{red}{2}, H: \textcolor{blue}{No}\} & \{A: \textcolor{red}{No}, M-H: \textcolor{red}{2}, M: \textcolor{red}{NA}, P: \textcolor{red}{1}, H: \textcolor{blue}{No}\} & \{A: \textcolor{red}{No}, M-H: \textcolor{red}{1}, H: \textcolor{blue}{No}\} & \{A: \textcolor{red}{No}, M-H: \textcolor{red}{1}, H: \textcolor{blue}{No}\}\\ 
    \cline{2-5}
    
    & \{A: \textcolor{red}{No}, M-H: \textcolor{red}{1}, 
    M: \textcolor{red}{NA}, P: \textcolor{red}{1}, H: \textcolor{blue}{No}\} & \{A: \textcolor{red}{No}, M-H: \textcolor{blue}{3}, M: \textcolor{blue}{Deny}, P: \textcolor{red}{2}, H: \textcolor{blue}{No}\} & \{A: \textcolor{red}{No}, M-H: \textcolor{red}{1}, H: \textcolor{blue}{No}\} & \{A: \textcolor{red}{No}, M-H: \textcolor{red}{1}, H: \textcolor{red}{Yes}\} \\  \hline

       \textbf{Social Causes and Stances} & Support Climate Action  & Deny Gender Equality & - & -  \\  \hline
   \textbf{Models} & MemeCraft-LLaMA & MemeCraft-LLaVA & Dank Learning & Random Online  \\  \hline
  \end{tabular}
  \caption{
 Example memes generated using various models with their  Authenticity (A), Meme Hilarity (M-H), Message Conveyance (M), Persuasiveness (P), and Hatefulness (H) scores. For clarity in our results, annotations that meet all the following criteria — Authenticity: Yes, Hilarity $\geq$ 3, Message Conveyance is correctly identified, Persuasiveness $\geq$ 3 and  Hatefulness: No — are highlighted in \textcolor{blue}{blue}. Conversely, annotations that do not meet these standards are highlighted in \textcolor{red}{red}. }
  
   \label{tab:case_study}
\end{table*}

\subsection{Case Study}
Table~\ref{tab:case_study} presents selected high-quality and low-quality memes across various models. Analysis of these case studies indicates that high-quality memes typically demonstrate a seamless integration of text and imagery, which together produce humor and a coherent message. A prime example is meme 2, where Drake's popular reaction image aptly satirizes the rejection of outdated gender stereotypes.

In contrast, low-quality memes can be attributed to several factors. Firstly, there may be a lack of semantic connection between the text and the image, as observed in memes 5 and 6. While the text may possess individual meaning, its relationship with the accompanying visual element is unclear. Secondly, some low-quality memes suffer from nonsensical textual content, as exemplified by meme 7. These memes fail to convey coherent messages, diminishing their overall quality. Finally, certain memes addressing sensitive topics may carry a latent risk of being labeled as hateful, as evidenced in the case of meme 8, which appears to downplay white people's historical involvement in the enslavement of Black individuals, potentially undermining their overall credibility.



\section{Conclusion}

In conclusion, our development of the \textsf{MemeCraft} generator constitutes a noteworthy advancement in automated meme creation. The research presented herein illustrates that by utilizing LLMs and VLMs, which have been pre-trained on comprehensive datasets, our system is capable of producing memes that not only closely mirror those found within online communities but also maintain contextual relevance and are devoid of offensive content. The generator has achieved satisfactory metrics in terms of persuasiveness and hilarity, indicating the utility of memes as a medium for supporting distinct societal objectives. While there is still room for improvement in achieving humor scores comparable to real online memes, our framework has significantly narrowed the gap. For future works, we aim to refine the synergy between textual and visual elements to enhance both the comedic and persuasive qualities of the generated memes, thereby contributing to the progressive evolution of meme generation technology.

\bibliographystyle{ACM-Reference-Format}
\bibliography{ref}

\begin{appendices}
\newpage


\section{Additional Information on Evaluation Results}
\begin{table*}[t]
\small
  \caption{Average authenticity scores of various models breakdown by persuasion techniques.}
  \begin{tabular}{ccccccccc}
    \toprule
    Social Cause &Model&Causes&Consequences&Solutions&Evidence of Absence&Benefits&Rationale&Average\\
    \midrule
    \multirow{2}{*}{Climate Action} 
    &  \textit{MemeCraft}-ChatGPT & 0.56& 0.47& 0.52& 0.41& 0.43&  -& 0.48\\
     & \textit{MemeCraft}-LLaMA& 0.5& 0.4& 0.41& 0.4& 0.45&  -& 0.43 \\
    & \textit{MemeCraft}-LLaVA & 0.41& 0.36& 0.32& 0.35& 0.36&  -& 0.36 \\
    \midrule
    \multirow{2}{*}{Gender Equality} &
    \textit{MemeCraft}-ChatGPT & 0.58& 0.58& 0.48& 0.56& -  & 0.47& 0.53  \\
     & \textit{MemeCraft}-LLaMA & 0.57& 0.47& 0.44& 0.34& -  & 0.45& 0.45  \\
    & \textit{MemeCraft}-LLaVA & 0.42& 0.33& 0.33& 0.28& -  & 0.32& 0.34  \\
     \midrule
     - & Dank Learning & - & - & - & -& - & - & 0.23  \\
     \midrule
    - & Online Random & - & - & - & -& - & - & 0.59 \\
  \bottomrule
\end{tabular}
\label{tab:similarity_appendix}
\end{table*}

\begin{figure*}[t]
  \centering
  \includegraphics[width=\textwidth]{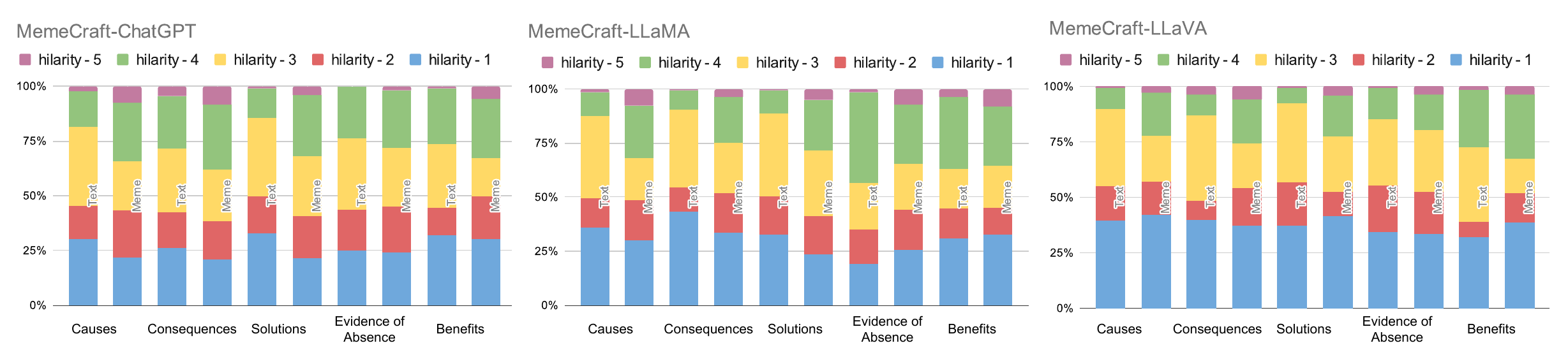}
  \caption{Distribution of hilarity scores across models for climate action breakdown by persuasion techniques.}
  \label{fig:hilarity_climate}
\end{figure*}

\begin{figure*}[t]
  \centering
  \includegraphics[width=\textwidth]{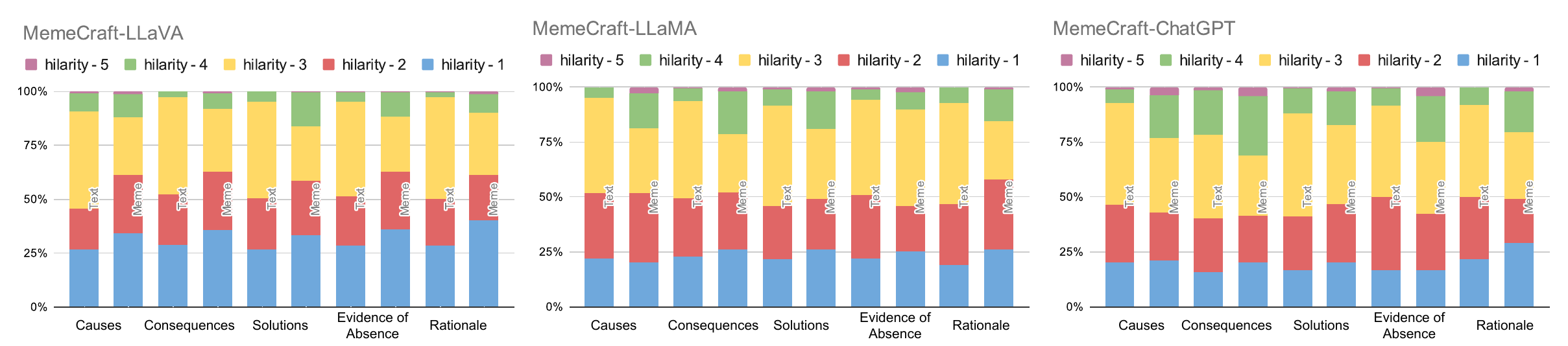}
  \caption{Distribution of hilarity scores across models for gender equality breakdown by persuasion techniques}
  \label{fig:hilarity_gender}
\end{figure*}

\begin{table*}
\small
  \caption{Hatefulness scores of generated memes identified
by the safety mechanism (Machine) breakdown by persuasion techniques.}
  \label{tab:freq}
  \begin{tabular}{ccccccccc}
    \toprule
    Social Cause &Model&Causes&Consequences&Solutions&Evidence of Absence&Benefits&Rationale&Average\\
    \midrule
    \multirow{3}{*}{Climate Action} &  \textit{MemeCraft}-ChatGPT & 0.03& 0.06& 0.07& 0.07& 0.04& -& 0.06\\
     & \textit{MemeCraft}-LLaMA& 0.01& 0.02& 0.05& 0.0& 0.08& -& 0.03 \\
    & \textit{MemeCraft}-LLaVA & 0.01& 0.02& 0.03& 0.07& 0.03& -&  0.03 \\
    \midrule
    \multirow{3}{*}{Gender Equality} &  \textit{MemeCraft}-ChatGPT & 0.54& 0.41& 0.56& 0.59& - & 0.63 & 0.55  \\
     & \textit{MemeCraft}-LLaMA & 0.38& 0.32& 0.39& 0.47& - & 0.5& 0.41 \\
    & \textit{MemeCraft}-LLaVA & 0.22& 0.28& 0.29& 0.28& - & 0.26 & 0.27 \\
  \bottomrule
\end{tabular}
\label{tab:hatefulness_model_appendix}
\end{table*}

\begin{table*}
\small
  \caption{Hatefulness scores of post-filtered memes
assess by the evaluators (Human) breakdown by persuasion techniques.}
  \label{tab:freq}
  \begin{tabular}{ccccccccc}
    \toprule
    Social Cause &Model&Causes&Consequences&Solutions&Evidence of Absence&Benefits&Rationale&Average\\
    \midrule
    \multirow{3}{*}{Climate Action} &  \textit{MemeCraft}-ChatGPT &  0& 0.03& 0.01& 0.02& 0.01& -& 0.01\\
     & \textit{MemeCraft}-LLaMA&  0& 0& 0.01& 0.01& 0.01& -& 0 \\
    & \textit{MemeCraft}-LLaVA &  0.01& 0.01& 0& 0.01& 0& -& 0.01 \\
    \midrule
    \multirow{3}{*}{Gender Equality} &  \textit{MemeCraft}-ChatGPT &  0.02& 0.02& 0.01& 0.01& -& 0.04& 0.02 \\
     & \textit{MemeCraft}-LLaMA &  0.03& 0.02& 0.03& 0.03& -& 0.08& 0.04 \\
    & \textit{MemeCraft}-LLaVA &  0.02& 0.01& 0.02& 0.01& -& 0.03& 0.01 \\
     \midrule
     - & Dank Learning & - & - & - & -& - & - & 0.05  \\
     \midrule
    - & Online Random & - & - & - & -& - & - & 0.05 \\
    
  \bottomrule
\end{tabular}
\label{tab:hatefulness_human_appendix}
\end{table*}

\end{appendices}

\end{document}